\documentclass[twocolumn,showpacs,preprintnumbers,amsmath,amssymb,10pt]{revtex4-1}
\usepackage{CJK}
\usepackage{graphicx}
\usepackage{subfigure}
\usepackage{dcolumn}
\usepackage{bm}
\usepackage{subeqnarray}
\usepackage{amsmath}
\usepackage{color}

\newcommand{\Fig}[1]{Fig.~\ref{#1}}

\begin{document}

\begin{CJK*}{GBK}{Song}

\title{Two-color narrowband photon pair source with high brightness \\ 
based on clustering in a monolithic waveguide resonator}

\date{\today}
\author{Kai-Hong Luo, Harald Herrmann, Stephan Krapick, Raimund Ricken,\\ Viktor Quiring, Hubertus Suche, Wolfgang Sohler, and Christine Silberhorn}

\affiliation{Integrated Quantum Optics, Applied Physics, University of Paderborn,
Warburger Str. 100, D-33098, Paderborn, Germany}

\begin{abstract}
We report on an integrated non-degenerate narrowband photon pair source produced at 890 nm and 1320nm via type II parametric down-conversion in a periodically poled waveguide with high-reflective dielectric mirrors deposited on the waveguide end faces. The conversion spectrum consists of three clusters and only 3 to 4 longitudinal modes with about 150 MHz bandwidth in each cluster. The high conversion efficiency in the waveguide, together with the spectral clustering in the double resonator, leads to a high brightness of $3\times10^3~$pairs/(s$\cdot$mW$\cdot$MHz). The compact and rugged monolithic design makes the source a versatile device for various applications in quantum communication.
\end{abstract}

%
%
\maketitle
\end{CJK*}

Single photon pair sources with narrow bandwidth and long coherence time
play an essential role in applications of quantum information processing; in particular for
quantum key distribution \cite{GisinRMP2002}, long distance quantum communication \cite{DuanN2001},
and the realization of quantum networks \cite{KimbleN2008}. To overcome current limitations of
long distance quantum communication due to transmission losses, quantum repeater architectures have been proposed \cite{BriegelPRL1998,SimonPRL2007,SangouardRMP2011}. These typically require photon pairs with one wavelength in the telecommunication band and one wavelength that matches the absorption line of the storage medium in a quantum memory (QM) \cite{LvovskyNP2009,WalmsleyNP2010,LamNP2011}. Such QMs usually have their absorption line in the visible or near infrared, i.e.\ far away from the telecommunication range, and the bandwidth is typically in the range of a few to several 100 MHz. Among the most promising materials for high-bandwidth QM's are solid-state atomic ensembles, specifically rare-earth ion doped  crystals or glasses \cite{TittelLPR2010,TittelN2011}. For example, a
Nd$^{3+}$-doped Y$_2$SiO$_5$ crystal has been successfully applied as an multi-mode atomic frequency comb memory with an absorption bandwidth of about 120 MHz \cite{GisinN2011}.

A well-known method to generate photon pairs is parametric down conversion
(PDC) \cite{HarrisPRL1967,BurnhamPRL1970}. In such a process, a medium with with $\chi^{(2)}$ nonlinearity splits a single pump photon into two photons of lower energy, named the signal and the idler, obeying with energy conservation and phase matching.
However, the loose phase-matching condition usually leads to a broad bandwidth typically exceeding several 100 GHz. As a result,
a strong filtering is required to match the acceptance bandwidth of the QMs with the drawback of losing most of the generated photon pairs. To overcome this bottle-neck, narrowband photon-pair sources are desired with an adapted bandwidth and a high spectral brightness.

One promising approach to generate such narrowband photon pairs is to
use resonance enhancement of PDC within a cavity, also called
optical parametric oscillator (OPO) far below the threshold \cite{OuPRL1999,WangPRA2004,
KuklewiczPRL2006, BensonPRL2009, PolzikOL2009, PanPRL2008, GuoOL2008,
WolfgrammPRL2011, PomaricoNJP2009, PomaricoNJP2012,
URenLP2010, ChuuAPL2012, FortschNC2013, FeketePRL2013}. PDC is enhanced at the resonances of the cavity but inhibited at non-resonant frequencies. In this way the spectral density is redistributed in comparison to the non-resonant case and, thus, ideally a filter-free source with actively reduced bandwidth and without sacrificing the photon flux level can be realized \cite{OuPRL1999}.

In most of the cavity-enhanced PDC experiments reported so far, the photon pairs were frequency degenerate or close to
frequency degeneracy. Thus, within the phase-matching bandwidth a comb of narrow lines is generated with a line spacing corresponding to the free spectral range (FSR) of the resonator. In the non-degenerate case, however, the material dispersion results in different FSRs for signal and idler. As maximum enhancement is only obtained if both signal and idler are resonant simultaneously, PDC is generated only in certain regions of the spectrum, so called 'clusters' \cite{EckardtJOSAB1991}. Within each cluster PDC occurs  only at a few longitudinal modes.  In this year the clustering approach has been applied for the first time to generate photon pairs with bandwidth of 2 MHz within 3 clusters (spaced 44 GHz) and 4 longitudinal modes inside of each cluster using a bulk crystal cavity etalon \cite{FeketePRL2013}.

Although a monolithic, resonant PDC source exploiting whispering gallery modes
has been presented recently \cite{FortschNC2013}, most of the
demonstrated resonant PDC sources used bulk crystals (either PPLN or PPKTP)
as the nonlinear material placed in an external resonator.
However, it is well-known that PDC in a waveguide is typically 2 to 3 orders of magnitude more efficient than in bulk crystals \cite{TanzilliEP2002,FiorentinoOE2007} due to the strong confinement of the optical fields along the whole interaction length. Obviously,  the resonator approach can be transferred to waveguide structures. Moreover, if the resonator is formed by mirrors directly deposited on the waveguide end-faces, a compact and rugged design requiring no cavity alignment can be realized. Tuning of the resonance frequency can be accomplished by varying the optical path length in the waveguide, for instance by temperature tuning \cite{SchreiberSPIE2001}.

Initial experiments with a resonant PPLN waveguide source based on degenerate type I phase-matching have been demonstrated in \cite{PomaricoNJP2009}. A detailed theoretical study revealed that using type II phase-matching should offer a reduced number of longitudinal modes \cite{PomaricoNJP2012}. This is due to the fact that the large birefringence in this system provides a larger difference between the FSR's of the signal and the idler fields.
Moreover, in this study it was pointed out that an optimized performance of such a resonant source must carefully take into consideration waveguide losses and resonator outcoupling efficiency to select design parameters like waveguide length or mirror reflectivities.

\begin{figure}[tbp]
\includegraphics*[width=0.4\textwidth]{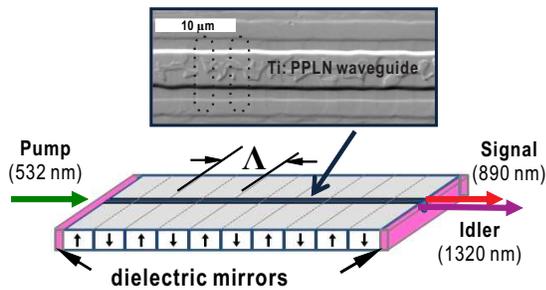}
\caption{(color online)  Integrated narrowband photon pair source composed of a Ti indiffused waveguide in a PPLN substrate with $\Lambda=4.44~\mu$m poling period and dielectric mirrors with high reflectivities at both signal and idler wavelengths deposited on the waveguide end-faces. The upper inset is a zoom to domain structure of the periodically poled waveguide area, and the black dotted lines denote the inverted domains.
\label{ResonantSample}}
\end{figure}

In this Letter, we present the first experimental realization of such an integrated compact photon pair source based on a doubly resonant waveguide exploiting type II phase-matching. The detailed structure of the integrated waveguide chip is shown in \Fig{ResonantSample}. The source consists of a 14.5~mm long Ti-indiffused waveguide in Z-cut PPLN. The poling period of
$\Lambda=$4.44~$\mu$m was chosen to provide first-order type II phase-matched PDC to generate TE-polarized signal photons around 890~nm and TM-polarized photons around 1320~nm when pumped at $\lambda_p=$532~nm in TE-polarization. This wavelength combination was chosen to fit to the Nd-based QM \cite{GisinN2011}.

Waveguide losses have been measured to be $\alpha_s \approx 0.05~$dB/cm and $\alpha_i \approx 0.06~$dB/cm for signal and idler, respectively. To implement the resonant source dielectric layers composed of alternating layer stacks of SiO$_2$ and TiO$_2$ were deposited on the end-faces of the waveguide. Based on the design rules given in \cite{PomaricoNJP2012} the resonator was modeled with a finesse exceeding 20 in order to provide the spectral narrowing. This could be best realized with an
asymmetric resonator with high reflectivities for signal and idler at the front mirror and reflectivities around 90~\% for the rear mirror (outcoupling mirror). With this asymmetry of the mirror reflectivities, the ratio of outcoupled signal (idler) photons to lost photons before escaping this cavity is about $\eta_s \approx 0.41$ ($\eta_i \approx 0.37$), respectively. Thus, the overall photon pair escape probability is given as ${\eta _{pp}} = {\eta _s}{\eta _i}$, which means about 15\% of the generated photon pairs leave  the cavity as couples at the desired output mirror.

In practice, a stack with 17 layers deposited as front mirror provides a reflectivity of $R\approx99\%$ for both wavelengths and the rear mirror consisting of 13 layers has the targeted $R\approx90\%$. The reflectivities of front and rear mirrors at 532~nm are 38\% and 8\%, respectively, to enable efficient incoupling of pump and to prevent triple resonance effects.
After mirror deposition the resonator was characterized carefully by launching appropriate narrowband light into the waveguide and monitoring the transmission while varying the cavity length via temperature tuning. We obtained a finesse of  $\mathcal{F}=22$ and $\mathcal{F}=25$ for the signal and idler wavelengths, respectively.

\begin{figure}[tbp]
\includegraphics*[width=0.49\textwidth]{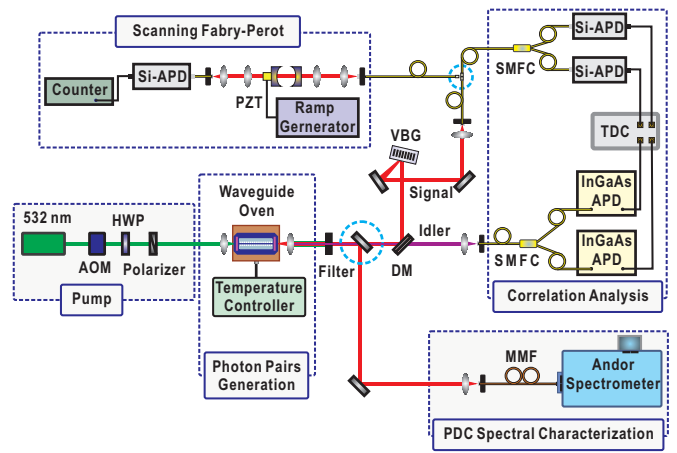}
\caption{(color online) Experimental setup for the generation and characterization of
the photon pairs from the resonant waveguide,
combining PDC characterization and coincidence measurements.
The detailed structure of the waveguide resonator chip in the oven is shown in \Fig{ResonantSample}.
The cyan dashed circle means the optical beam paths are alternative.
AOM: acousto-optical modulator; HWP: half-wave plate; DM: dichroic mirror;
SMFC: single-mode fiber coupler; MMF: multi-mode fiber; VBG: volume Bragg grating;
PZT: piezoelement; APD: avalanche photodiode; TDC: time-to-digital converter.}
\label{QuRePSetup}
\end{figure}

The experimental setup to study the resonant source is shown in
Fig.~\ref{QuRePSetup}. The sample is always pumped with a laser at 532 nm
with a specified bandwidth of less than 1 MHz.  To avoid the effects of photorefraction,
pump pulses with a typical
length of about 200 ns and a repetition rate of about 100 kHz are extracted from
the cw-source by using an acousto-optical modulator (AOM). By using a half wave plate (HWP)
together with a polarizer, the pump power is tunable from 0.1~mW to 10~mW.
The sample is heated to temperatures around 160~$^{\circ}$C to
obtain quasi-phase-matching for the desired wavelength
combination and to prevent luminescence and deterioration due to photorefraction. During the
measurements the sample temperature is stabilized to about $\pm$1 mK.

\begin{figure}[tbp]
\includegraphics*[width=0.4\textwidth]{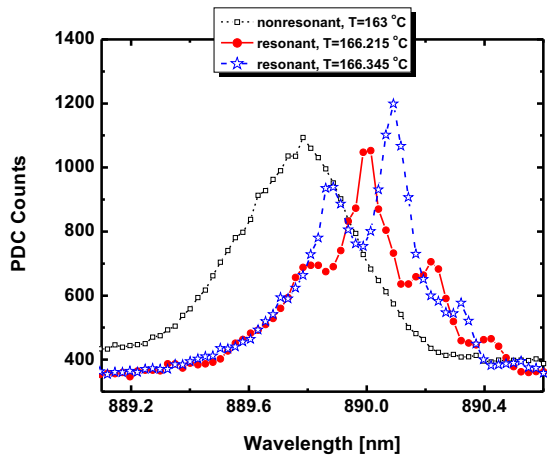}
\caption{(color online) PDC spectra measured using the uncoated waveguide (black square)
and the resonant waveguide at various temperatures (blue star, red circle).
The background counts can be attributed to the dark noise of CCD camera.}
\label{NarrowPDCenhance}
\end{figure}

To characterize the generated PDC in the signal wavelength range the output
from the waveguide is coupled to a spectrometer system (Andor Shamrock 303i with iKon-M 934 CCD camera)  with a resolution of
about 0.15 nm. In \Fig{NarrowPDCenhance} recorded PDC spectra of one waveguide prior and after mirror deposition
are shown. The spectrum of the uncoated waveguide shows a main peak with
a bandwidth (FWHM) of 0.4 nm ($\approx$ 151 GHz), as predicted for the 14.5 mm long interaction length. The spectra
of the same waveguide with dielectric mirrors, however, shows a pronounced sub-structure.
Although the resolution of the spectrometer is
too coarse to reveal details of the spectra, one can already derive that the pair generation occurs within
three clusters with a cluster separation of about 0.2 nm ($\approx$ 75 GHz). This separation is determined by the difference
of the FSRs of signal and idler, which theoretically are FSR$_s\sim4.4$~GHz and FSR$_i\sim4.7$~GHz \cite{PomaricoNJP2012,Luo2013}.
The cluster separation is half of the bandwidth of the PDC envelope. Ideally, there is a dominant
cluster in the the central of normalized phase-matching, and another two symmetric clusters
at the side wings of the envelope with about 41\% intensity.
Thus, three clusters can be observed within this envelope
with a strong temperature dependence as verified by comparing the two spectra shown in \Fig{NarrowPDCenhance} with only 30 mK temperature difference.

Moreover, these measurements are already a proof of resonance enhancement as can be easily seen by
comparing the spectra of the same waveguide before coating and after coating shown in \Fig{NarrowPDCenhance}.
For the measurements, all of the parameters, apart from the temperature, have been kept constant. In particular, the pump power about 1~mW in front of the incoupling lens has not been changed. If the cavity of the resonant sample only act as a narrowband filter, the overall level of the PDC measured with the spectrometer should have dropped significantly, caused by the integration over 0.15 nm, which is performed by the spectrometer due to its limited resolution. However, as the detected power levels are almost identical for the resonant and the non-resonant sample, we can conclude that due to the cavity there is a
spectral density redistribution resulting in an enhanced PDC generation at the cavity resonances.

To investigate details of the spectra of the resonant source we studied the internal structure
within a single cluster.  A volume Bragg grating (VBG, OptiGrate 900) with a spectral bandwidth
of 0.17 nm is inserted into the signal beam path as shown in \Fig{QuRePSetup} to act as bandpass filter to select a single cluster.
The filtered light is routed via a single mode fiber to a scanning
confocal Fabry-P\'{e}rot resonator with 15 GHz FSR and a finesse of about 20. Its transmission is analyzed using
a single photon detection module (Perkin Elmer Avalanche photodiode SPCM-AQR-14)
to record the signal photons transmitted from the Fabry-P\'{e}rot  which is  tuned by applying a voltage ramp to the piezo-driven
mirror mount.

\begin{figure}[tbp]
\includegraphics*[width=0.4\textwidth]{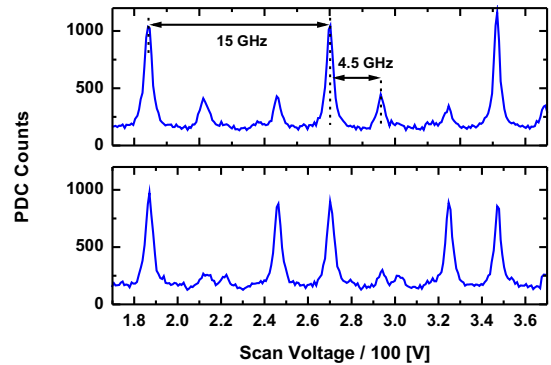}
\caption{(color online) Signal spectra recorded with a
confocal scanning Fabry-P\'{e}rot with 15 GHz FSR.
The two spectra vary in temperature by 6mK around 161.57~$^{\circ}$C.}
\label{NarrowPDCSFP}
\end{figure}

Corresponding measurement results are shown in Fig.~\ref{NarrowPDCSFP}, they reveal
the modal structure within a single cluster.  The spectrum consists of longitudinal modes with
4.5~GHz frequency separation. By finely tuning the temperature, we are able to suppress the two 'satellite' modes, leaving a predominantly single mode operation -- as in the upper case of \Fig{NarrowPDCSFP}. A slight shift of the temperature by less than 6~mK
leads to a spectrum composed of two modes with almost equal strength surrounded by weak additional satellite modes.
The measured linewidth of each longitudinal mode of about 750 MHz is mainly determined by the resolution of the scanning
Fabry-P\'{e}rot resonator, but it is not the natural bandwidth of the generated PDC photons.

\begin{figure}[tbp]
\includegraphics*[width=0.4\textwidth]{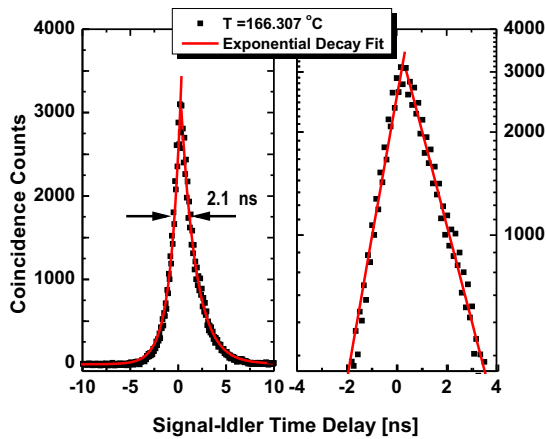}
\caption{(color online) Measured coincidences of photon pairs
as a function of arrival time difference between
the signal and idler photons using linear scaling (left) and logarithmic scaling (right) with exponential fits.}
\label{NarrowG2CoinSI}
\end{figure}

To investigate the spectral linewidth of the photons and the correlation between photon pairs, coincidences between signal and idler have been characterized by measuring the arrival times of the respective photons with the set-up shown in Fig.~\ref{QuRePSetup}.  In the left diagram of \Fig{NarrowG2CoinSI}, the results from such a measurement are shown. It reveals a correlation time (FWHM of the coincidence peak) of about 2.1~ns. This is significantly broader than the corresponding results obtained with non-resonant samples showing a width of about 0.5~ns, which is determined by the finite resolution of our measurements system, due to timing jitters of the detection system. The correlation time is inversely proportional to the bandwidth of the down-conversion fields.
From the measured correlation time $\tau_{coh}$ of 2.1~ns, according to $\tau_{coh}=1/{\pi \Delta\nu}$, where $\Delta\nu$ is the bandwidth of the down-converted photons, a spectral bandwidth of about 150~MHz can be deduced. This is in good accordance with the theoretically predicted width of the resonances calculated for the given cavity parameters.

The presence of the cavity implies that the signal and idler photons may be emitted at distinct times, corresponding to a different number of round-trips within the cavity. Thus, the shape of the coincidence curve should be determined by exponential functions.
In the right diagram of \Fig{NarrowG2CoinSI} the measurement result is re-drawn using a logarithmic scaling together with exponential fits for the rising and falling parts. The slight asymmetry reflects the different finesses of signal and idler resulting in slightly different leakage times out of the resonator.

\begin{figure}[tbp]
\includegraphics*[width=0.4\textwidth]{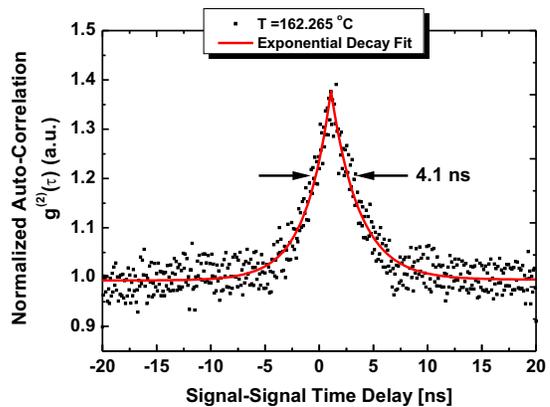}
\caption{(color online) Measured signal-signal autocorrelation as function of arrival time difference between
two signal photons. }
\label{NarrowG2CoinSS}
\end{figure}

Such coincidence measurements can also be evaluated to determine the efficiency of the PDC generation process \cite{Klyshko1980}. From the ratio of the coincidence to single counts the generated photon pair rate can be determined. We found that this generation rate is almost equal for both the non-resonant and resonant waveguides. In the resonant case, however, the spectral distribution is strongly confined to only a few longitudinal modes, whereas in the non-resonant case it is
distributed over the entire phase-matching bandwidth. For our source we determined a normalized generation rate (inside the resonator) of about $7\times10^6~$pairs/(s$\cdot$mW). Assuming these are distributed over three inequivalent clusters with three longitudinal modes with different weights within each cluster as shown in Fig.~\ref{NarrowPDCSFP}, we can estimate that about 39\% of the generated photon pairs are within the most dominant longitudinal mode with 150 MHz bandwidth. Taking into account the photon pair escape probability of  15~\%, the spectral brightness can be estimated to be  $B = 3\times10^3~$pairs/(s$\cdot$mW$\cdot$MHz).

An alternative method to characterize the spectral properties of the source is the analysis of the auto-correlation Glauber function $g^{(2)}(\tau)$ \cite{Glauber1963}. The number of effective modes K can be estimated from the normalized auto-correlation function at zero time delay according to  $g^{(2)}(0)=1 + 1/ K$ \cite{McNeilPRA1983,ChristNJP2011}.
We have measured the signal-signal autocorrelation by splitting  the signal radiation behind  behind the VBG (i.e.\ a single cluster was selected) using a 50:50 single mode fiber coupler (SMFC) and routing the two output ports of the SMFC to individual silicon single photon detectors.

An example of a  measured $g^{(2)}(\tau)$ characteristic is shown in \Fig{NarrowG2CoinSS}. Around zero time delay
$g^{(2)}$ strongly exceeds 1 proving the generation of non-classical light. It is shown that from the measured data $g^{(2)}(0)\approx 1.4$. As a result, an effective mode number of filtered signal beam $K=2.5$ can be estimated. On the other hand, the auto-correlation is only constant for the idler unfiltered beam. This is in reasonably good qualitative agreement with the measured spectra shown in \Fig{NarrowPDCSFP}, where we observed three modes (with different amplitudes) within a single cluster. A refined theoretical model would be necessary to study quantitatively the relationship between $g^{(2)}(\tau)$  and the observed spectra.

In summary, we have experimentally demonstrated a compact, bright and narrowband photon pair source by exploiting clustering in a doubly resonant Ti:PPLN waveguide. In the type II phase-matched, nondegenerate PDC source photon pairs
are generated with one photon matching the absorption line of a Nd-based QM, and the wavelength of the idler photon (around 1320 nm) is compatible for long distance transmission over standard telecom fibers.
Although the measured bandwidth of about 150 MHz is still about two orders of magnitude larger than the bandwidth demonstrated with bulk optical versions \cite{FeketePRL2013}, the compact and rugged design and the large brightness of about $3\times10^3~$pairs/(s$\cdot$mW$\cdot$MHz) makes this device a versatile source for various quantum applications. In particular, it has a great potential if an ultimate narrow bandwidth is not required. Beyond further fundamental studies and work on technological improvements, future activities can focus on the development of pure, filter-free tunable single mode sources using pulsed light. One can exploit the use of the cluster effect together with a double-pass pumping scheme to restrict PDC generation to only a single cluster \cite{ChuuAPL2012}. Further improvement of the resonator quality might enable the restriction to, ultimately, a single longitudinal mode, which can further be tuned via a monolithicly integrated electro-optic modulator.

The authors thank Benjamin Brecht and Malte Avenhaus for helpful discussions. We gratefully acknowledge the support by the
European Union through the QuReP project (no. 247743).


\begin{thebibliography}{99}

\bibitem{GisinRMP2002} N. Gisin, G. Ribordy, W. Tittel, and H. Zbinden, Rev. Mod.
Phys. \textbf{74}, 145 (2002).

\bibitem{DuanN2001} L. M. Duan, M. D. Lukin, J. I. Cirac, and P. Zoller,
Nature \textbf{414}, 413 (2001).

\bibitem{KimbleN2008} H. J. Kimble, Nature \textbf{453}, 1023 (2008).

\bibitem{BriegelPRL1998} H.-J. Briegel, W. D\"{u}r, J. I. Cirac, and P. Zoller,
Phys. Rev. Lett. \textbf{81}, 5932 (1998).

\bibitem{SimonPRL2007} C. Simon, H. de Riedmatten, M. Afzelius, N. Sangouard,
H. Zbinden, and N. Gisin, Phys. Rev. Lett. \textbf{98}, 190503 (2007).

\bibitem{SangouardRMP2011} N. Sangouard, C. Simon, H. de Riedmatten, and N. Gisin, Rev. Mod.
Phys. \textbf{83}, 33 (2011).

\bibitem{LvovskyNP2009}
A. I. Lvovsky, B. C. Sanders, and W. Tittel, Nat. Photon. \textbf{3}, 706 (2009).

\bibitem{WalmsleyNP2010}
K. F. Reim, J. Nunn, V. O. Lorenz, B. J. Sussman, K. C. Lee, N. K. Langford,
D. Jaksch, and I. A. Walmsley, Nat. Photon. \textbf{4}, 218 (2010).

\bibitem{LamNP2011}
M. Hosseini, G. Campbell, B. M. Sparkes, P. K. Lam, B. C. Buchler,
Nat. Phys. \textbf{7}, 794 (2011).

\bibitem{TittelLPR2010}
W. Tittel, M. Afzelius, T. Chaneli\'{e}re, R. L. Cone, S. Kr\"{o}ll,
S. A. Moiseev, and M. Sellars, Laser \& Photon. Rev. \textbf{4}, 244. (2010).

\bibitem{TittelN2011}
E. Saglamyurek, N. Sinclair, J. Jin, J. A. Slater, D. Oblak, F. Bussi\`{e}res,
M. George, R. Ricken, W. Sohler, and W. Tittel, Nature, \textbf{469}, 512 (2011).

\bibitem{GisinN2011} C. Clausen, I. Usmani, F. Bussi\`{e}res, N. Sangouard,
M. Afzelius, H. de Riedmatten, and N. Gisin, Nature \textbf{469}, 508, (2011).

\bibitem{HarrisPRL1967} S. E. Harris, M. K. Oshman, and R. L. Byer, Phys. Rev.
Lett. \textbf{18}, 732 (1967).

\bibitem{BurnhamPRL1970} D. C. Burnham and D. L. Weinberg, Phys. Rev. Lett. \textbf{25}, 84 (1970).

\bibitem{OuPRL1999} Z. Y. Ou and Y. J. Lu, Phys. Rev. Lett. \textbf{83}, 2556 (1999).

\bibitem{WangPRA2004} H. Wang, T. Horikiri, and T. Kobayashi, Phys. Rev. A \textbf{70}, 043804 (2004).

\bibitem{KuklewiczPRL2006} C. E. Kuklewicz, F. N. C. Wong, and J. H. Shapiro, Phys.
Rev. Lett. \textbf{97}, 223601 (2006).


\bibitem{BensonPRL2009} M. Scholz, L. Koch, and O. Benson,
Phys. Rev. Lett. \textbf{102}, 063603 (2009).

\bibitem{PolzikOL2009} B. Melholt Nielsen, J. S. Neergaard-Nielsen, and E. S. Polzik,
Opt. Lett. \textbf{34}, 3872 (2009).

\bibitem{PanPRL2008} X.-H. Bao, Y. Qian, J. Yang, H. Zhang, Z.-B. Chen, T. Yang,
and J.-W. Pan, Phys. Rev. Lett. \textbf{101}, 190501 (2008).

\bibitem{GuoOL2008} F.-Y. Wang, B.-S. Shi, and G.-C. Guo, Opt. Lett. \textbf{33},
2191 (2008).


\bibitem{WolfgrammPRL2011} F. Wolfgramm, Y. A. de Icaza Astiz, F. A. Beduini, A. Cer\`{e}, and M. W. Mitchell,
Phys. Rev. Lett. \textbf{106}, 053602 (2011).


\bibitem{PomaricoNJP2009} E. Pomarico, B. Sanguinetti, N. Gisin, R. Thew, H. Zbinden,
G. Schreiber, A. Thomas and W. Sohler, New J. Phys. \textbf{11}, 113042 (2009).

\bibitem{PomaricoNJP2012} E. Pomarico, B. Sanguinetti, C. I. Osorio, H. Herrmann
and R. T. Thew, New J. Phys. \textbf{14}, 033008 (2012).

\bibitem{URenLP2010} Y. Jeronimo-Moreno, S. Rodriguez-Benavides, and
A. B. U'Ren, Laser Phys. \textbf{20}, 1221 (2010).

\bibitem{ChuuAPL2012} C.-S. Chuu, G. Y. Yin, and S. E. Harris,
Appl. Phys. Lett. \textbf{101}, 051108 (2012).

\bibitem{FortschNC2013} M. F\"{o}rtsch, J. F\"{u}rst, C. Wittmann, D. Strekalov, A. Aiello,
M. V. Chekhova, C. Silberhorn, G. Leuchs, and C. Marquardt,  Nat. Commun. \textbf{4}, 1818 (2013).

\bibitem{FeketePRL2013} J. Fekete, D. Riel\"{a}nder, M. Cristiani, H. de Riedmatten,
Phys. Rev. Lett. \textbf{110}, 220502 (2013).

\bibitem{EckardtJOSAB1991} R. C. Eckardt, C. D. Nabors, W. J. Kozlovsky, and R. L. Byer, J. Opt. Soc.
Am. B \textbf{8}, 646 (1991).

\bibitem{TanzilliEP2002}
S. Tanzilli, W. Tittel, H. De Riedmatten, H. Zbinden, P. Baldi, M. De Micheli,
D. B. Ostrowsky, and N. Gisin, Eur. Phys. J. D \textbf{18}, 155 (2002).

\bibitem{FiorentinoOE2007}
M. Fiorentino, S. M. Spillane, R. G. Beausoleil, T. D. Roberts, P. Battle,
and M. W. Munro, Opt. Expr. \textbf{15}, 7479 (2007).

\bibitem{SchreiberSPIE2001}
G. Schreiber, D. Hofmann, W. Grundk\"{o}ter, Y. L. Lee,
H. Suche, V. Quiring, R. Ricken and W. Sohler, Proc. SPIE \textbf{4277}, 144 (2001).

\bibitem{Luo2013}
K.-H. Luo {\sl et al.}, in preparation .

\bibitem{Klyshko1980}
D. N. Klyshko, Sov. J. Quantum Electron \textbf{10}, 1112, (1980).

\bibitem{McNeilPRA1983} K. J. McNeil, C. W. Gardiner, Phys. Rev. A - General
Physics \textbf{28}, 1560 (1983).

\bibitem{Glauber1963} R. J. Glauber, Phys. Rev. \textbf{131}, 2766 (1963).

\bibitem{ChristNJP2011} A. Christ, K. Laiho, A. Eckstein, K. N. Cassemiro, and C. Silberhorn,
New J. Phys. \textbf{13}, 033027, (2011).

\end{thebibliography}
\end{document}